\documentstyle[sprocl,cite,epsfig]{article}

\begin{document}

\title{
\mbox{}\hfill{\rm
\vbox{\hbox{SNUTP-00-009}\hbox{nucl-th/0004055}}}\\[1cm]
ELECTROMAGNETIC PRODUCTION OF VECTOR MESONS AT LOW ENERGIES%
\footnote{Talk at the NSTAR2000 Workshop, {\it The Physics of Excited
Nucleons\/}, JLab, Newport News, Feb. 16--19, 2000}}

\author{Yongseok OH}

\address{Center for Theoretical Physics, Seoul National University,
Seoul 151-742, Korea \\ and \\
Institute of Physics, Academia Sinica, Nankang, Taipei, Taiwan 11529,
R.O.C. \\ E-mail: ohys@phys.sinica.edu.tw}

\author{Alexander I. TITOV}

\address{Bogoliubov Laboratory of Theoretical Physics, JINR, Dubna
141980, Russia \\ E-mail: atitov@thsun1.jinr.ru}

\author{T.-S. Harry LEE}

\address{Physics Division, Argonne National Laboratory, Argonne,
Illinois 60439,
U.S.A. \\ E-mail: lee@anph09.phy.anl.gov}


\maketitle\abstracts{ 
We have investigated exclusive photoproduction of light vector mesons
($\omega$, $\rho$ and $\phi$) on the nucleon at low energies.
In order to explore the questions concerning the so-called missing
nucleon resonances, we first establish the predictions from a model
based on the Pomeron and meson exchange mechanisms.
We have also explored the contributions due to the mechanisms
involving $s$- and $u$-channel intermediate nucleon state.
Some discrepancies found at the energies near threshold and large
scattering angles suggest a possibility of using this reaction
to identify the nucleon resonances.
}

At high energies and low momentum transfers, the exclusive
electromagnetic production of vector mesons has been explained
successfully by the Pomeron exchange model. \cite{DL84,LM95,PL97}
However, at low energies near threshold, meson exchange mechanisms 
become important, such as the $\pi$ exchange in $\omega$ production and
the $\sigma$ exchange in $\rho$ production. \cite{FS96}
Furthermore, the mechanisms involving intermediate nucleon and nucleon
resonances ($N^*$), which could be suppressed at high energies, must
also be included in a complete theoretical investigation.

To resolve the so-called ``missing resonance problem'', it is essential to
identify the kinematic regions where the $N^*$ contributions are important.
Many of those missing nucleon resonances are predicted to have large
partial widths for their decays into channels consisting of the nucleon
and a vector meson.
Therefore, one may hope to find them in vector meson electromagnetic
productions.
There exist some reports on the contributions from the nucleon resonances
in vector meson photoproduction based on quark models \cite{ZLB98c,ZDGS99}
or the Regge theory \cite{Lage00}.
However, in order to confront the forthcoming data, it is crucial to
understand the non-resonant (background) production processes which could
interfere strongly with the resonant production amplitudes.
As a step in this direction, we have investigated the $\rho$, $\omega$ and
$\phi$ photoproductions at low energies based on a model consisting of 
three mechanisms: Pomeron exchange, one-meson ($\pi$, $\eta$, $\sigma$)
exchange and the mechanism involving an intermediate nucleon in $s$-channel
and $u$-channel (called $s+u$ nucleon-term from now on).

\begin{figure}
\centering
\epsfig{file=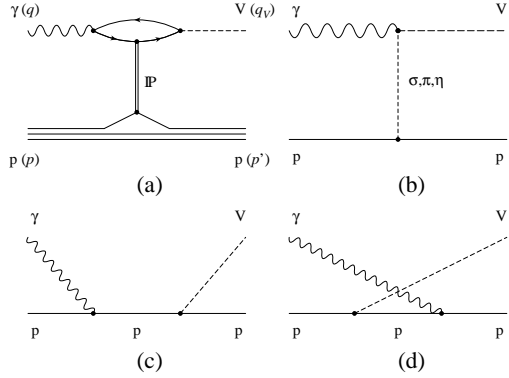,height=2.0in}
\caption{Three mechanisms for vector meson $(V = \rho, \omega, \phi$)
photoproduction: (a) Pomeron, (b) one-boson exchange,
(c) and (d) $s$- and $u$-channel intermediate nucleon diagrams.}
\label{fig:vm}
\end{figure}

Each of the considered production amplitude, as illustrated in
Fig.~\ref{fig:vm}, can be written as 
\begin{equation}
T_{fi} = \varepsilon_\mu^*(V) {\cal M}^{\mu\nu}
\varepsilon_\nu^{}(\gamma),
\label{Tmat}
\end{equation}
where $\varepsilon_\mu^{}(V)$ and $\varepsilon_\nu^{}(\gamma)$ are the
polarization vectors of the vector meson and the photon, respectively.
We first consider the Pomeron exchange depicted in Fig.~\ref{fig:vm}(a).
In this process, the incoming photon first converts into a $q\bar{q}$
pair, which interacts with the nucleon by the Pomeron exchange before
forming the outgoing vector meson.
The quark-Pomeron vertex is obtained by the Pomeron-photon analogy,
\cite{DL84} which treats the Pomeron as a $C=+1$ isoscalar photon, as
suggested by a study of nonperturbative two-gluon exchanges. \cite{LN87}
We then have \cite{DL84,LM95,PL97,TOYM98}
\begin{eqnarray}
{\cal M}^{\mu\nu}_P &=&
i 12 \sqrt{4\pi\alpha_{\rm em}} \beta_u G_P (w^2,t)
F_1(t) \frac{m_V^2 \beta_f}{f_V}
\frac{1}{m_V^2 - t}
\left( \frac{2\mu_0^2}{2 \mu_0^2 + m_V^2 - t} \right)
\nonumber \\ && \mbox{} \times
\bar{u}_{m'} (p') \bigl\{
\not \! q \, g^{\mu\nu} - q^\mu \gamma^\nu
\bigr\} u_m(p),
\label{Tpom}
\end{eqnarray}
where $\alpha_{\rm em} = e^2/4\pi$, $m$ and $m'$ are the 
spin projections of the initial and final nucleons, respectively.
Here we denote the four-momenta of the initial nucleon, final nucleon,
incoming photon and outgoing vector meson by $p$, $p'$, $q$ and
$q^{}_V$, respectively.
Their helicities are represented by $\lambda_p^{}$, $\lambda_p'$,
$\lambda_\gamma^{}$ and $\lambda_V^{}$.
The Mandelstam variables are $s = W^2 = (p+q)^2$, $t = (p-p')^2$,
$u = (p-q_V^{})^2$.
The proton and vector meson masses are represented by $m_p$ and $m_V^{}$,
respectively, and $F_1$ is the isoscalar electromagnetic form factor of
the nucleon,
\begin{equation}
F_1(t) = \frac{4m_p^2 - 2.8t}{(4m_p^2 - t)(1-t/0.71)^2}.
\end{equation}
The Pomeron-exchange is described by the following Regge form,
\begin{equation}
G_P (w^2,t) = \left( \frac{w^2}{s_0^{}} \right)^{\alpha_P^{} (t) - 1}
\exp\left\{ - \frac{i\pi}{2} [ \alpha_P^{} (t) - 1 ] \right\},
\end{equation}
with $w^2 = (2W^2 + 2 m_p^2 - m_V^2)/4$ and $s_0^{} = 1/\alpha_P'$.
The Pomeron trajectory is taken to be the usual form 
$\alpha_P^{} (t) = 1.08 + \alpha'_P t$ with $\alpha'_P = 0.25$
GeV$^{-2}$. 
In Eq. (2), $f_V$ is the vector meson decay constant: $f_\rho = 5.04$,
$f_\omega = 17.05$ and $f_\phi = 13.13$.
The coupling constants $\beta_u=\beta_d = 2.07$ GeV$^{-1}$, 
$\beta_s = 1.60$ GeV$^{-1}$ and $\mu_0^2 = 1.1$ GeV$^2$ are chosen to
reproduce the total cross section data at high energies $E_\gamma \ge 10$
GeV where the vector meson photoproductions are completely dominated
by Pomeron-exchange.

For the one-meson exchange diagram of Fig.~\ref{fig:vm}(b), we consider
scalar and pseudoscalar meson exchanges.
The vector meson exchange is not allowed in this process and the possible
exchange of axial vector mesons\cite{KMOV00} is suppressed at low
energies mainly because of their heavy masses and small coupling constants.

\begin{table}[t]
\centering
\begin{tabular}{|c|c|c|c|} \hline
  & $\rho$  & $\omega$  & $\phi$ \\ \hline
$g_{V\gamma\pi}^{}$ & $0.274$ & $0.706$ & $0.042$ \\ \hline
$g_{V\gamma\eta}^{}$ & ---    & $0.062$ & $0.209$ \\ \hline
\end{tabular}
\caption{Coupling constants $g^{}_{V\gamma\pi}$ and $g^{}_{V\gamma\eta}$
in unit of GeV$^{-1}$.}
\label{tab:Vgp}
\end{table}

The pseudoscalar meson exchange amplitude can be obtained from
the Lagrangian,
\begin{equation}
{\cal L}_\varphi = g^{}_{V\gamma\varphi} \epsilon^{\mu\nu\alpha\beta}
\partial_\mu V^{}_\nu \partial_\alpha A_\beta \varphi
 -i g^{}_{\varphi NN} \bar N \gamma_5 \varphi N,
\end{equation}
where $\varphi = (\pi^0, \eta)$ and $A_\mu$ is the photon field.
The coupling constants $g^{}_{V\gamma\pi}$ and $g^{}_{V\gamma\eta}$,
as given in Table~\ref{tab:Vgp}, are obtained from the experimental
partial widths \cite{PDG98} of the vector meson radiative decays
$V \rightarrow \gamma \varphi$. 
We use $g^2_{\pi NN} / 4\pi = 14.0$ and the SU(3) relation to
obtain $g_{\eta NN} / g_{\pi NN} \simeq 0.35$.
To account for the effects due to the finite hadron size at each vertex,
the resulting Feynman amplitudes are regularized by the following form
factors,
\begin{equation}
F_{\varphi NN} = \frac{\Lambda_\varphi^2 - M_\varphi^2}
                {\Lambda_\varphi^2 - t }, \qquad
F_{V\gamma\varphi} = \frac{\Lambda_{V\gamma\varphi}^2 - M_\varphi^2}
{\Lambda_{V\gamma\varphi}^2 - t },
\end{equation}
where ($\Lambda_\pi = 0.7$, $\Lambda_{V\gamma\pi} = 0.77$) and
($\Lambda_\eta = 1.0$, $\Lambda_{V\gamma\eta} = 0.9$) in GeV unit.
\cite{TLTS99}

The scalar ($\sigma$) meson exchange was introduced in Ref. \cite{FS96}
to describe the $\rho$ photoproduction.
This can be considered as an effective way to account for the two-$\pi$
exchange in $\rho$ production, which is expected to be significant because
of the large branching ratio of $\rho \to \pi^+ \pi^- \gamma$ decay.
The contribution from the $\sigma$ exchange can be obtained from
the following Lagrangian,
\begin{equation}
{\cal L}_\sigma = \frac{eg^{}_{V\gamma\sigma}}{M_V^{}} \left( \partial^\mu
V^\nu \partial_\mu A_\nu - \partial^\mu V^\nu \partial_\nu A_\mu \right)
\sigma +
g_{\sigma NN} \bar{N} N \sigma.
\end{equation}
We also regularize the resulting one-meson-exchange amplitude by
the form factors,
\begin{equation}
F_{\sigma NN} = \frac{\Lambda_\sigma^2 - M_\sigma^2}
                {\Lambda_\sigma^2 - t }, \qquad
F_{V\gamma\sigma} = \frac{\Lambda_{V\gamma\sigma}^2 - M_\sigma^2}
{\Lambda_{V\gamma\sigma}^2 - t }.
\end{equation}
Following Ref. \cite{FS96}, we use $M_\sigma = 0.5$ GeV,
$g^2_{\sigma NN} / 4\pi = 8.0$, $\Lambda_\sigma = 1.0$ GeV and
$\Lambda_{V\gamma\sigma} = 0.9$ GeV.
The coupling constant $g_{V\gamma\sigma}^{}$ is around $3.0$ for
reproducing the total cross sections of $\rho$ photoproduction near
threshold.
We do not consider the $\sigma$-exchange in $\omega$ and $\phi$
photoproductions, since the radiative decays of these two vector mesons
into $\pi^+\pi^-$ states are much weaker than that into single pion state.
This could be understood by considering the current-field identity.
\cite{FS96} (See also Ref. \cite{TLTS99}.)

Finally we consider the $s+u$ nucleon-term shown in Fig.~\ref{fig:vm}(c,d).
The corresponding amplitudes can be obtained from the following
Lagrangian,
\begin{equation}
{\cal L}_{N}^{} =
g_{V NN}^{} \bar{N} \left[ \gamma_\mu V^\mu
- \frac{\kappa_V^{}}{2 M_N^{}} \sigma_{\mu\nu} \partial^\nu V^\mu
  \right] N
-e  \bar{N} \left[ \gamma_\mu A^\mu
- \frac{\kappa_p^{}}{2 M_N^{}} \sigma_{\mu\nu} \partial^\nu A^\mu
  \right] N,
\end{equation}
where the anomalous magnetic moment of the proton is $\kappa_p^{} = 1.79$.
The coupling constants are chosen to be $(g_{\rho NN}^{}, \kappa_\rho^{})
= (6.2, 2.0)$ and  $(g_{\omega NN}^{}, \kappa_\omega^{}) = (7.0, 0)$,
as determined in a study of $\pi N$ scattering and pion
photoproduction. \cite{SL96} 
In this study, we do not consider the $s+u$ nucleon-term in $\phi$
photoproduction by assuming $g^{}_{\phi NN} \approx 0$ due to the OZI rule.
A possible modification due to nonvanishing $\phi NN$ coupling can be
found in Ref.~\cite{TLTS99}.
The form factor $F_V^{}$ for $VNN$ vertex is assumed to be of the form
given by Ref. \cite{HBMF98a},
\begin{equation}
F_V (s,u) =
\frac12 \left( \frac{\Lambda_{VNN}^4}{\Lambda_{VNN}^4 + (s - M_N^2)^2}
+ \frac{\Lambda_{VNN}^4}{\Lambda_{VNN}^4 + (u - M_N^2)^2}
\right),
\end{equation}
with $\Lambda_{VNN} = 0.8$ GeV.


\begin{figure}[t]
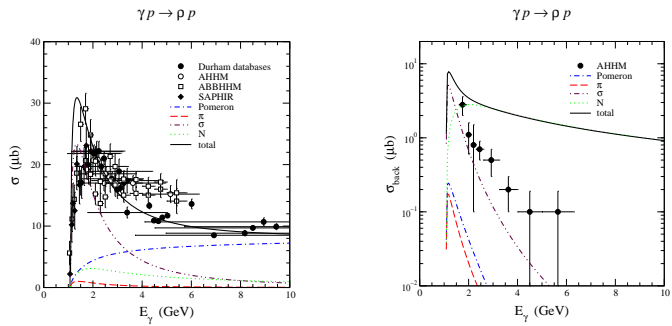

\centering
\epsfig{file=rho_total_cs.eps,height=1.5in, angle=-90} \qquad\quad
\epsfig{file=rho_back_cs.eps,height=1.5in, angle=-90}
\caption{(Left figure) total cross section of $\rho$ photoproduction. The
dot-dashed, dashed, dot-dot-dashed, dotted and solid lines are Pomeron
exchange, $\pi$ exchange, $\sigma$ exchange, intermediate nucleon and
the sum of the amplitudes, respectively. The experimental data are from
Refs. \cite{Durham,AHHM76,ABBH68,Klein96-98}.
(Right figure) cross section $\sigma_{\rm back}$ with
the data from Ref.~\cite{AHHM76}.}
\label{fig:rho_cs}
\end{figure}

\begin{figure}
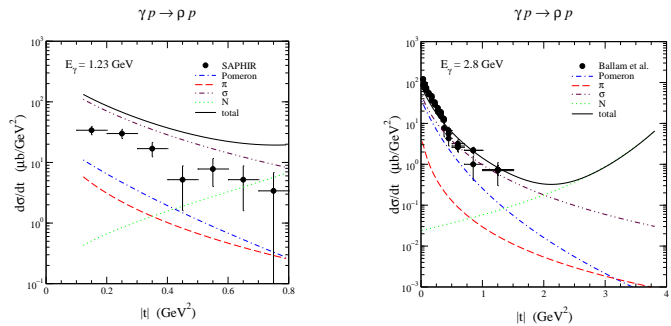

\centering
\epsfig{file=rho_dsdt_1_23.eps,height=1.5in, angle=-90} \qquad\quad
\epsfig{file=rho_dsdt_2_8.eps,height=1.5in, angle=-90}
\caption{Differential cross sections of $\rho$ photoproduction at
$E_\gamma = 1.23$ and $2.8$ GeV.
The experimental data are from Refs.~\cite{Klein96-98,BCGG72BCEK73}.
Notations are the same as in Fig.~\ref{fig:rho_cs}.}
\label{fig:rho_dsdt}
\end{figure}

With the production amplitudes given above, we have explored the extent
to which the existing data of vector meson photoproductions can be
described by the non-resonant (background) mechanisms.
The $\rho$ photoproduction cross sections are calculated from the
amplitudes due to Pomeron, $\pi$, $\sigma$ exchanges and the $s+u$
nucleon-term .
In the left panel of Fig.~\ref{fig:rho_cs} we show that the calculated
total cross sections (solid curve) agree to a very large extent with the
data up to $E_\gamma = 10$ GeV.
The important role of the $\sigma$ exchange (dot-dot-dashed curve) at
energies near threshold is also shown there. 
The dynamical content of our model can be better seen by investigating
angular distributions.
For example, given in the right panel of Fig.~\ref{fig:rho_cs} is the
backward integrated cross sections defined by
\begin{equation}
\sigma_{\rm back}^{} \equiv \int_{\pi/2}^{\pi} \frac{d\sigma}{d\theta}
d\theta.
\end{equation}
The predicted backward cross sections (solid curve) clearly overestimate
the experimental data.
We find that this is caused by the dominant contributions (dotted curve)
from the $s+u$ nucleon-term .
On the other hand, the calculated cross sections will be too low if these
mechanisms involving an intermediate nucleon state are not included in
the calculation.
This suggests that some other production mechanisms, which can give
important contributions at large scattering angles, must be included in
a more complete model. 
This can also be seen in the angular distributions of the differential
cross sections shown in Fig.~\ref{fig:rho_dsdt}.
We see that the contribution from the $s+u$ nucleon-term (dotted curves)
become dominant at large scattering angles. 
Therefore, precise measurements of the differential cross sections in
the backward scattering region will shed light on the role of the
intermediate nucleon states.
At very low energies near threshold, $E_\gamma \le 2$ GeV, our predictions
overestimate significantly the data.
This is perhaps mainly due to the neglect of the $N^*$ excitations in our
calculations.

\begin{figure}[t]
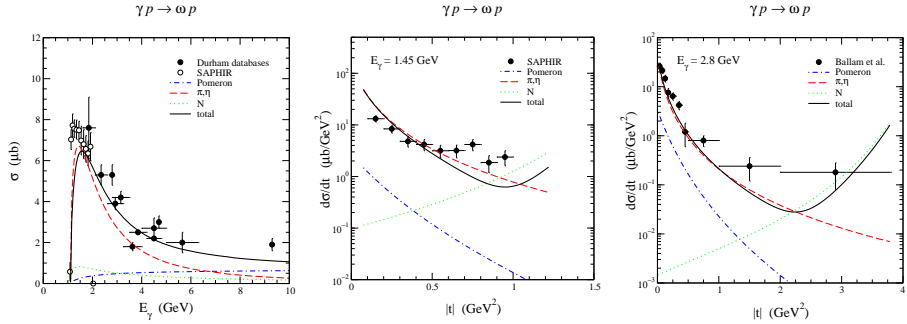

\centering
\epsfig{file=omega_total_cs.eps,height=1.5in, angle=-90} \hfill
\epsfig{file=om_dsdt_1_45.eps,height=1.5in, angle=-90} \hfill
\epsfig{file=om_dsdt_2_8.eps,height=1.5in, angle=-90}
\caption{Total and differential cross sections of $\omega$ photoproduction.
The experimental data are from Refs. \cite{Durham,Klein96-98,BCGG72BCEK73}.
Notations are the same as in Fig.~\ref{fig:rho_cs}.}
\label{fig:omega_cs}
\end{figure}

The $\omega$ photoproduction cross sections are calculated from the
Pomeron, $\pi$ and $\eta$ exchanges and the $s+u$ nucleon-term.
The predicted total and differential cross sections are compared with
the data in Fig.~\ref{fig:omega_cs}.
We see that the one-pion exchange (long dashed curves) dominates $\omega$
photoproduction up to $E_\gamma \approx 6$ GeV.
The predicted total cross sections somewhat underestimate the recent
SAPHIR data in the $E_\gamma \le 2$ GeV region where the mechanisms
involving the excitation of nucleon resonances are expected to play
some roles.
Similar to the case of $\rho$ photoproduction, the $s+u$ nucleon-term
also dominates the cross sections at large $|t|$ region.

\begin{figure}[t]
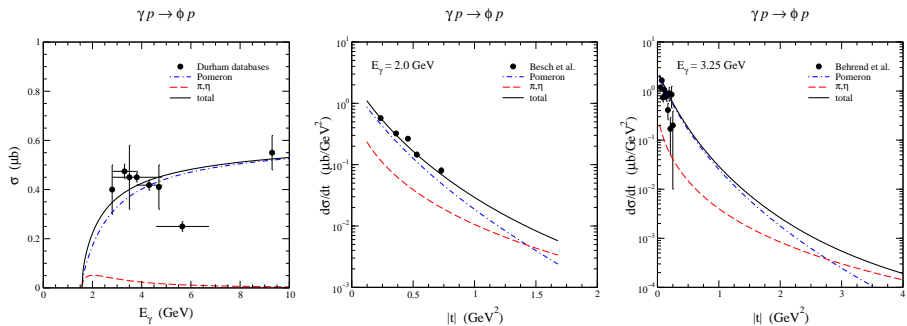

\centering
\epsfig{file=phi_total_cs.eps,height=1.5in, angle=-90} \hfill
\epsfig{file=phi_dsdt_2_0.eps,height=1.5in, angle=-90} \hfill
\epsfig{file=phi_dsdt_3_25.eps,height=1.5in, angle=-90} 
\caption{Total and differential cross sections of $\phi$ photoproduction.
The experimental data are from Refs. \cite{Durham,BHKK74,BBHM78}.
Notations are the same as in Fig.~\ref{fig:rho_cs}.}
\label{fig:phi_cs}
\end{figure}

For $\phi$ photoproduction, we consider the Pomeron and pseudoscalar
meson exchanges only.
We refer the calculations on the scalar meson exchanges and direct
$\phi$ radiation arising from the non-vanishing $\phi NN$ coupling
to Ref.~\cite{TLTS99} and the effects of nonvanishing strangeness of
the nucleon to Refs.~\cite{TOY97,TOYM98}.
In Fig.~\ref{fig:phi_cs} we show that the predicted total and
differential cross sections agree well with the very limited data.
Contrary to the cases of $\rho$ and $\omega$ photoproductions, the
Pomeron exchange (dash-dotted curves) gives the major contribution to
the total cross section even at energies near threshold. 
This is mainly due to the fact that the coupling of $\phi$ to the nucleon
is suppressed by the OZI rule.
The contributions from $\pi$ and $\eta$ exchanges are also found to be
small.
Therefore $\phi$ photoproduction can provide a useful tool to study the
nature of the Pomeron exchange.

In summary, we have investigated the exclusive photoproduction of light
vector mesons ($\rho$, $\omega$ and $\phi$) at low energies.
Our model includes the Pomeron exchange, meson ($\pi$, $\eta$, $\sigma$)
exchange and the $s$- and $u$-channel nucleon terms.
It is found that the predicted cross sections agree with the existing data
to a large extent.
However, some significant discrepancies are found in the total and
differential cross sections at low energies and at large scattering
angles.
It is possible that such discrepancies could be removed by extending our
model to include the effects due to the excitation of nucleon resonances.
Our investigation in this direction is in progress, aiming at using the
new data from current experimental facilities to test various QCD-inspired
models of hadron structure.


\section*{Acknowledgments}
We are grateful to V.~Burkert, W.-C. Chang, F.~J. Klein, N.~I. Kochelev
and T.~Nakano for fruitful discussions and informations.
Y.O. thanks the Theory Division of Argonne National Laboratory and the
organizer of this workshop for the warm hospitality and financial
support.
This work was supported in part by
KOSEF of Korea, NSC of Republic of China,
Russian Foundation for Basic Research under Grant No 96-15-96426
and U.S. DOE Nuclear Physics Division Contract No. W-31-109-ENG-38.


\vfill\mbox{}
\eject

\end{document}